\documentclass{ifacconf}
\usepackage{amsmath,amssymb,amsfonts}
\usepackage{graphicx}      % include this line if your document contains figures
\usepackage{natbib}        % required for bibliography
\begin{document}
\begin{frontmatter}

\title{Modelling and Synchronisation of Delayed Packet-Coupled Oscillators in Industrial Wireless Sensor Networks\thanksref{footnoteinfo}}

\thanks[footnoteinfo]{Yan Zong gratefully acknowledges financial support from the University of Northumbria at Newcastle \emph{via} a postgraduate research studentship.}
\thanks[footnoteinfo]{\copyright \ $2020$ Zong, Y. et al. This work has been accepted to IFAC for publication under a Creative Commons Licence CC-BY-NC-ND.}

\author{Yan Zong, }
\author{Xuewu Dai, }
\author{Pep Canyelles-Pericas, }
\author{Krishna Busawon, }
\author{Richard Binns, }
\author{Zhiwei Gao}

\address{Department of Mathematics, Physics and Electrical Engineering, Northumbria University, Newcastle upon Tyne, United Kingdom (e-mail: \{yan.zong, xuewu.dai, pep.canyelles-pericas, krishna.busawon, richard.binns, zhiwei.gao\}@northumbria.ac.uk).}

\begin{abstract}                % Abstract of not more than 250 words.
In this paper, a Packet-Coupled Oscillators (PkCOs) synchronisation protocol is proposed for time-sensitive Wireless Sensor Networks (WSNs) based on Pulse-Coupled Oscillators (PCO) in mathematical biology. The effects of delays on synchronisation performance are studied through mathematical modelling and analysis of packet exchange and processing delays. The delay compensation strategy (i.e., feedforward control) is utilised to cancel delays effectively. A simple scheduling function is provided with PkCOs to allocate the packet transmission event to a specified time slot, by configuring reference input of the system to a non-zero value, in order to minimise the possibility of packet collision in synchronised wireless networks. The rigorous theoretical proofs are provided to validate the convergence and stability of the proposed synchronisation scheme. Finally, the simulations and experiments examine the effectiveness of PkCOs with delay compensation and scheduling strategies. The experimental results also show that the proposed PkCOs algorithm can achieve synchronisation with the precision of $26.3\mu s$ ($1$ tick).
\end{abstract}


\end{frontmatter}
%===============================================================================

\section{Introduction}
Nowadays, the wide use of Internet of Things (IoT) in various sectors, such as transportation systems, electrical grids, smart cities, and so on, is significantly improving the quality of our life. As a fundamental technology of IoT, Wireless Sensor Networks (WSNs) plays an important role in industrial systems by monitoring and collecting surrounding context and environmental information. To guarantee such time-sensitive and mission-critical industrial applications work properly, Time Synchronisation (TS), which is the key enabling technology of WSNs, is required to provide a common timescale for local clocks of WSN nodes.

Due to the importance of time synchronisation, many synchronisation protocols have been proposed for WSNs in the communication engineering community. They are well represented by TS algorithms such as Precision Time Protocol (PTP) (\cite{ieee:1588}) and Reference Broadcast Synchronisation (RBS) of \cite{Elson:2002}. It is well known that in the PTP synchronisation scheme, error in each hop is accumulated, and the precision of such a synchronisation algorithm degrades by increasing hop distance. Compared to the PTP protocol, RBS achieves higher performance at the expense of requiring more packet transmission and reception, while the Radio Frequency (RF) module is the most energy-consuming component of a wireless sensor node, which would increase energy complexity (\cite{Sivrikaya:2004}).

Next, owing to the significance of synchronisation in various disciplines, it has naturally attracted attention from mathematics and physics communities, where it is applied to networked oscillators consisting of a set of oscillators coupled with each other through a certain network topology. Inspired by the synchronous flashing of fireflies observed in certain parts of southeast Asia, a typical networked oscillators model, the Pulse-Coupled Oscillators (PCO), is derived by \cite{Mirollo:1990}. As a result of its inherent simplicity and scalability, this biological model guarantees the possibility of synchronisation in a network consisting of $N$ oscillators under simple networking protocols; its discretisation also lets this model apply to wireless sensor networks easily.

However, the PCO's ideal coupling mechanism restricts its employment to any real network. Specifically, in PCO, oscillators' states are instantly known by their coupled oscillators and no delays between connected oscillators, which is impossible in industrial wireless networks. Thus, it needs to be improved for application in WSNs.

As a result, in this paper, inspired by PCO, the Packet-Coupled Oscillators (PkCOs) is proposed to achieve time synchronisation in WSNs. The main focus is particularly laid on the problem of the ideal coupling mechanism (i.e., packet exchange delay and processing delay in WSNs).

In PkCOs, an oscillator works in either free-running or interactive mode. In the free-running mode, the oscillator behaves as an isolated oscillator, whose state $P$ linearly increases from zero to the threshold. Once $P$ reaches the threshold value, the oscillator fires (which means a \emph{Sync} packet is being generated and broadcasted) while state $P$ is reset. This procedure keeps repeating continuously during the free-running mode. When the oscillator receives a \emph{Sync} from another oscillator, it moves to the interactive mode, where generating a local timestamp, and adjusting its state by offset estimated from the timestamp, after which it returns to the free-running mode.

Typically, in embedded systems, the clock module of a WSN node is implemented by a counter register (or a chain of counters) driven by the crystal oscillator. The counter is reset to zero when it matches the pre-defined threshold value. This periodic resetting feature is similar to PkCOs' firing-resetting behaviour, where the oscillator fires and resets its state. Thus, the clock module in embedded systems can be described as an uncoupled PkCOs' oscillator, and periodic packet transmission in WSNs is equivalent to \emph{Sync} firing in PkCOs.

\subsection{Related Work}
In PCO, the presence of pulse coupling delay results in the occurrence of infinite feedback and instability of networks (\cite{Hong:2005}). To prevent WSNs from being unstable, several works (e.g., \cite{Hong:2005}, \cite{Pagliari:2011}) introduce a refractory period where the reception of \emph{Pulse}s has no effect on local oscillator's state, and refractory period must be larger than twice coupling delay (\cite{Gentz:2016}, \cite{Tyrrell:2008}). However, \cite{Zong:2018a} shows that, in multi-hop networks, PCO synchronisation error (resulting from coupling delay) accumulates over multiple hops. Therefore, TS error may be larger than the refractory period, which is utilised to guarantee the network's stability. Moreover, due to the impossibility of real-time computing, the effects of processing delay (which is required for calculation) need to be considered in the synchronisation protocol.

In the literature (e.g., see \cite{Tyrrell:2010}, \cite{Proskurnikov:2017}, \cite{Wang:2012}, \cite{Hong:2005}, \cite{Pagliari:2011} and \cite{An:2011}), once a WSN consisting of numerous sensor nodes achieves synchronisation, a large number of \emph{Sync} packets are sent to the wireless channel simultaneously. However, each time only a single wireless packet is permitted to be transmitted in the channel. As a consequence, packet collision occurs, and no packets can be received successfully. To address this issue, \cite{Ashkiani:2012} and \cite{Gentz:2016} propose a scheduling scheme to guarantee \emph{Sync} packets to be transmitted in a distributed fashion, thereby minimising the possibility of packet collision. However, the dithered quantisation function, which is used for scheduling scheme, is not naturally integrated into WSN nodes. The microprocessor in sensor nodes has to provide extra hardware resources to realise dithered quantisation function, which also increases energy consumption, puts a strain in batteries, and reduces the autonomy of nodes. Furthermore, the use of software floating-point unit in dithered quantisation function may reduce its calculation precision.

\subsection{Contribution and Paper Organisation}
In this paper, through mathematical modelling of packet exchange and processing delays in both time dimension and state dimension, the delays' effects on PkCOs in these dimensions are analysed. Delay compensation strategy (i.e., feedforward control) is introduced in the packet coupling scheme to cancel the impacts of two delays, rather than simply introducing a refractory period to maintain the network's stability, which cannot improve synchronisation performance effectively. Also, the packet coupling mechanism allows the natural allocation of time slots for \emph{Sync} packet transmission, and the simple scheduling function of packet coupling mechanism reduces the need and consumption of hardware resources, compared to scheduling functions in \cite{Ashkiani:2012} and \cite{Gentz:2016}.

In addition, the convergence and stability of PkCOs are analysed under two different scenarios (i.e., with and without feedforward control). The effectiveness of the proposed delay compensation scheme is also evaluated in simulation and hardware-based experiments. The experimental results show that, by using the proposed synchronisation strategy, the time synchronisation precision of $26.3 \mu s$ ($1$ tick) is achieved under clock resolution of $30.5 \mu s$; and packet coupling scheme is also capable of allocating \emph{Sync} packet transmission to a specified time slot.

The rest of this paper is organised as follows: Section $2$ presents the PkCOs algorithm and mathematical modelling of delays in the time dimension. Section $3$ details the implementation of a delay compensation scheme in the state dimension. The rigorous theoretical proofs for the algorithm's convergence and stability are also provided. Additionally, the scheduling function is briefly introduced in this section. Section $4$ and Section $5$ present simulation and experimental results. Finally, conclusions are drawn in Section $6$.

\section{Problem Formulation}
In embedded systems, the crystal oscillator is widely used as a clock source due to its trade-off between high-quality signal accuracy and cost. In the following, the crystal oscillator-based clock is modelled as a free-running oscillator of PkCOs. An identical clock model is considered since the crystal oscillator provides a reliable clock signal. To understand how clock and delays are modelled, their equations are derived and briefly presented.

\subsection{Modelling An Identical Clock}
Referring first to the case of a perfect clock, in WSNs, the clock system is constructed from two parts: (i) a crystal oscillator, ticking at the nominal frequency $f_0$ = $1/\tau_0$ where $\tau_0$ is crystal oscillator period, and (ii) a counter counts the number of ticks generated by a crystal oscillator. Specifically, through the process of counting, the periodic signal produced by a crystal oscillator is converted into an integer that is increased by one per crystal oscillator period. Once the cumulative value of a counter matches the pre-defined threshold, it is reset and starts counting from zero again, after which this procedure repeats. Let $t$[$n$] denote the time reported by such a crystal oscillator-based clock at the $n$-th clock event, $t$[$n$] is calculated as
\begin{equation}
t[n] = n \times \tau_0 = \frac{n}{f_0}. \label{eq1}
\end{equation}
For an ideal crystal oscillator-based clock whose frequency is exactly the same as the nominal frequency $f_0$, $t$[$n$] is accurate and referred to as the \emph{reference time}. Such a clock is also called the \emph{reference clock} or \emph{master clock} thereafter.

In the PkCOs synchronisation method, the time synchronisation cycle $T$ is much greater than the clock period $\tau_0$ (i.e., $T \gg \tau_0$), it is reasonably assumed that the clock is updated $m_0$ times during a single synchronisation cycle, following $T = m_0 \times \tau_0$. Taking the clock's periodic resetting behaviour into account, the dynamics of the master clock's state $P_0$[$n$] are represented as
\begin{equation}
P_0[n]
=t[n] - \sum_{j=0}^{k}\varphi_0[j], \label{eq2}
\end{equation} where $\varphi_0$ is the master clock threshold, $k = \lfloor n/m_0 \rfloor$ represents how many clock resetting has occurred from $n = 0$ to the $n$-th clock event, where floor function $\lfloor n/m_0 \rfloor$ denotes the greatest integer less than or equal to $n/m_0$. Recalling that synchronising clocks occurs when a clock resets, $k$ also denotes the number of synchronisation cycles so far. In other words, the clock is at the $k$-th synchronisation cycle.

However, in reality, the state of real crystal oscillator-based clock $P_i$[$n$] is different from ideal state $P_0$[$n$] due to manufacturing tolerance and environmental conditions (e.g., temperature). The difference between ideal clock's state $P_0$[$n$] and $P_i$[$n$] reported by the $i$-th unperfect clock is referred to as the clock offset $\theta_i$[$n$]:
\begin{equation}
\theta_i[n] = P_i[n] - P_0[n]. \label{eq3}
\end{equation}
Clearly, in Fig. 1, the amount of offset $\theta_i$[$n$] of the crystal oscillator-based clock is moderate throughout experimental recordings at room temperature. By using the linear polynomial fitting, it is shown that the clock possesses a smaller clock skew $\gamma_i$[$n$]\footnote{clock skew $\gamma_i$[$n$] is defined as the normalised difference between the $i$-th clock frequency $f_i$[$n$] and \emph{reference clock} frequency $f_0$, yielding $\gamma_i[n]=(f_i[n]-f_0)/f_0$.}, which is only around $1.4$ ppm (parts per million). Thus, it is reasonable to assume that the frequency $f_i$[$n$] of the crystal oscillator-based clock is identical to nominal clock frequency $f_0$[$n$] (i.e., $\gamma_i$[$n$] = $0$). Thus, only the crystal oscillator's phase contributes to the inaccuracy of local clocks. The phase noise of the $i$-th clock is modelled by a random process $\phi_i$[$n$], representing all the instant phase deviation from $t$[$0$] to time $t$[$n$]. The clock state $P_i$[$n$] at the $n$-th clock event is modelled as
\begin{equation}
P_i[n]
=t[n] + \phi_i[n] -  \sum_{j=0}^{k}\varphi_i[j], \label{eq4}
\end{equation} where $\varphi_i$ is the $i$-th clock threshold. The local unperfect clock is also called the \emph{slave clock}. For the purpose of analysis, the recursive state equation of slave clock, which describes the behaviour of a clock with identical frequency, is written as
\begin{equation}
\theta_i[n+1] = \theta_i[n] + \omega_{\theta_i}[n], \label{eq5}
\end{equation} where white Gaussian random process $\omega_{\theta_i}[n] = \phi_i[n+1] - \phi_i[n]$ denotes offset noise.

\begin{figure}
\begin{center}
\includegraphics[width=8.4cm]{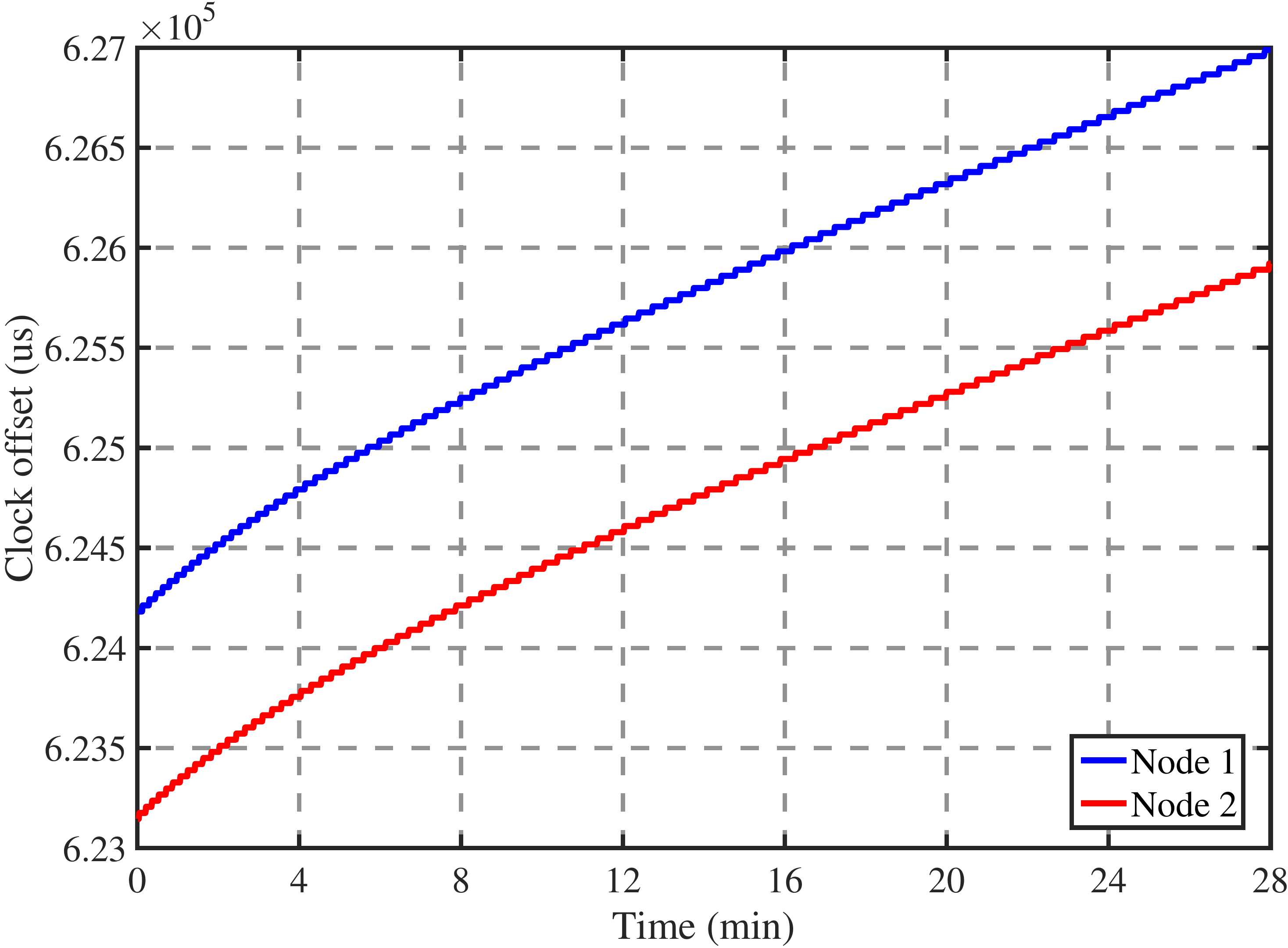}    % The printed column width is 8.4 cm.
\caption{Offsets of $32.768$kHz crystal oscillator-based clocks at wireless nodes (Atmel SMART SAM R21).}
\label{fig:fig1}
\end{center}
\end{figure}

\subsection{Mathematical Modelling of Delays}
In the PkCOs' packet coupling scheme, at the $k$-th synchronisation cycle, when the slave node receives the master's \emph{Sync} after packet exchange delay $\kappa_i$, it generates a timestamp $\hat{P}_i$[$k$] based on the local clock. While, due to the impossibility of real-time computing, the processing delay $\eta_i$ is required for a processor to estimate offset $\hat{\theta}_i$[$k$]. Once the offset estimate $\hat{\theta}_i$[$k$] is determined, the correction input $u_i$[$k$] is employed to drifting clock (5) instantly.

Obviously, the inaccuracy of clock correction action $u_i$[$k$] results from following two factors: (i) packet exchange delay $\kappa_i$ including send time, access time, propagation time and receive time; and (ii) processing delay $\eta_i$ which is required for a processor to compute clock correction input from the local timestamp. These factors are taken into accounts as shown in Fig. 2.

\begin{figure}
\begin{center}
\includegraphics[width=8.4cm]{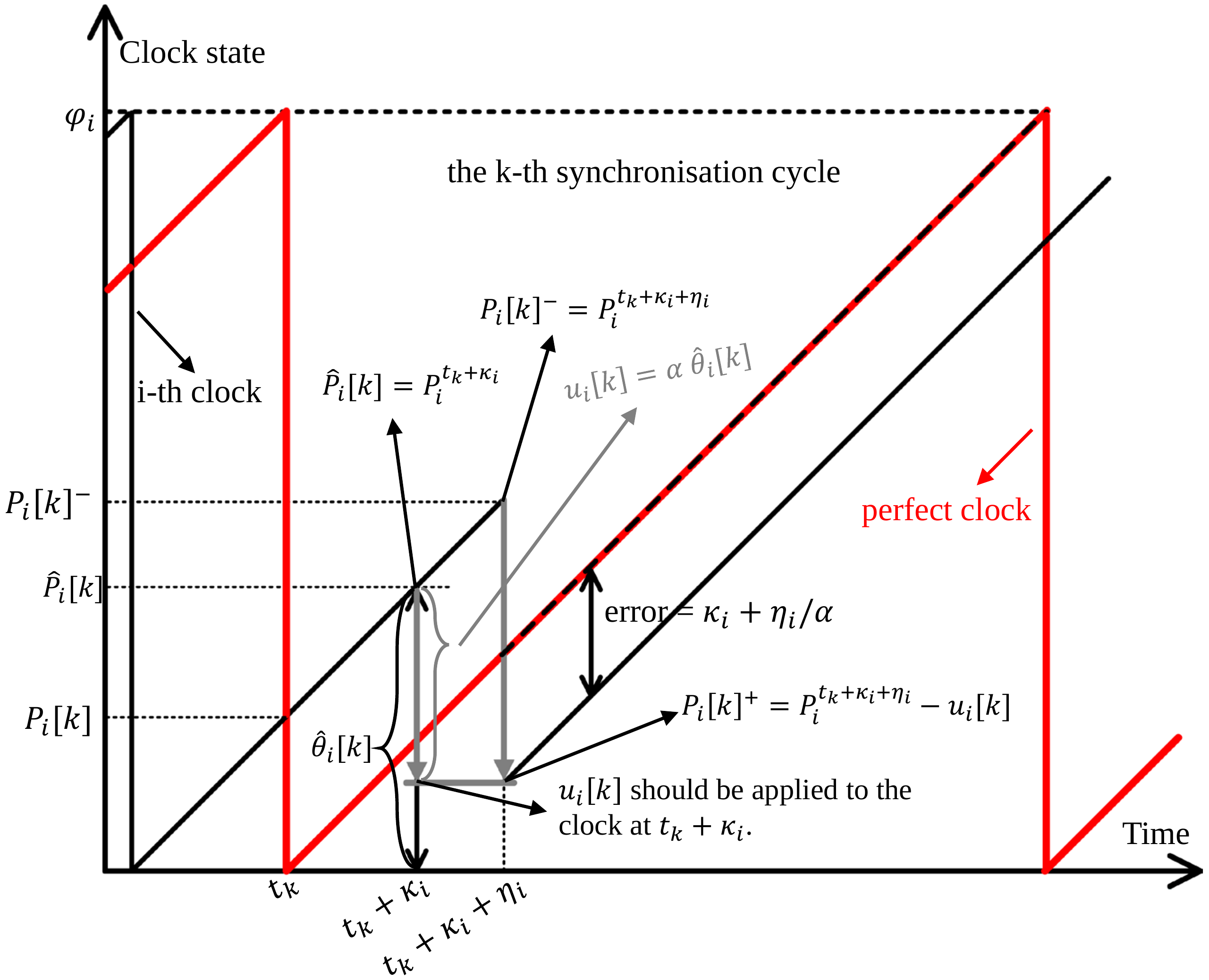}    % The printed column width is 8.4 cm.
\caption{Example of the slave clock correction.}
\label{fig:fig2}
\end{center}
\end{figure}

\emph{Packet Exchange Delay}: As shown in Fig. 2, the perfect clock fires and transmits a \emph{Sync} packet at $t_k$, which indicates the start of the $k$-th synchronisation cycle. The $i$-th node receives the \emph{Sync} at $t_k + \kappa_i$, due to the transmission delay, propagation delay, and other factors explained above. The packet exchange delay $\kappa_i$ is assumed as Gaussian random process with non-zero mean of $\bar{\kappa}_i$ and finite variance of $ \sigma_{\kappa_i}^2 $. That is $\kappa_i \sim (\bar{\kappa}, \sigma_{\kappa_i}^2 )$.

When the $i$-th node receives the \emph{Sync}, a local timestamp $P_i$[$n$] is generated by reading node's counter, we get:
\begin{equation}
\hat{P}_i[k] = \hat{P}_i^{t_k+\kappa_i}. \label{eq6}
\end{equation}
Once local timestamp $\hat{P}_i$[$k$] is obtained, the offset estimate $\hat{\theta}_i[k] = \theta_i[k]+\kappa_i$ between two clocks can be determined directly via the following expression:
\begin{equation}
\hat{\theta}_i[k] = \Bigg \{ \begin{array}{cc}
\hat{P}_i[k] & if \hat{P}_i[k]<\frac{\varphi_i}{2}+\kappa_i\\
\hat{P}_i[k] - \varphi_i & if \hat{P}_i[k]\geq\frac{\varphi_i}{2}+\kappa_i
\end{array}.\label{eq7}
\end{equation}
\emph{Processing Delay}: Next, the offset estimate should be employed to a local clock at $t_k + \kappa_i$:
\begin{equation}
P_i[k]^+ = P_i[k]^- - u_i[k], \label{eq8}
\end{equation}
where $P_i[k]^+/P_i[k]^-$ is the $i$-th clock's state after/before it is corrected at the $k$-th synchronisation cycle. However, due to the limitation of hardware resources, the processing delay $\eta_i$ is required for clock offset estimation. Thus, the local clock actually is corrected at the time $t_k + \kappa_i + \eta_i$, yielding
\begin{equation}
P_i[k]^+ = P_i^{t_k + \kappa_i + \eta_i} - u_i[k], \label{eq9}
\end{equation}
where the processing delay $\eta_i$ is also considered as Gaussian random process with non-zero mean $\bar{\eta}_i$ and finite variance of $ \sigma_{\eta_i}^2 $, which is $\eta_i \sim (\bar{\eta}, \sigma_{\eta_i}^2 )$.

In (3) and (4), it is indicated that the existence of clock offset leads to the clock state's inaccuracy. Therefore, the clock correction action to clock state is essentially equivalent to the application of correction input $u_i$[$k$] to clock offset. For the purpose of analysis, the clock correction algorithm (9) can be rewritten as
\begin{equation}
\theta_i[k]^+ = \theta^{t_k + \kappa_i + \eta_i}_i - u_i[k], \label{eq10}
\end{equation}
where $\theta_i[k]^+$ is the clock offset after it is adjusted.

Until now, it is clear that due to the packet exchange delay and processing delay in the time dimension, the direct employment of complete offset estimate to a local clock, yielding $u_i[k] = - \hat{\theta}_i[k]$, results in degradation of PkCOs synchronisation performance.

\section{Packet Coupling Scheme With Feedforward Control}
To avoid the occurrence of over-correction and synchronisation performance's degradation, this paper applies a proportional controller scheme for better synchronisation performance. Moreover, by effectively compensating the effects of delays in the state dimension, the feedforward control is capable of further improving synchronisation precision.

% \begin{figure}
% \begin{center}
% \includegraphics[width=8.4cm]{fig3}    % The printed column width is 8.4 cm.
% \caption{Block diagrams of the closed-loop control system (a) the closed-loop control system with delays in the time dimension, (b) the closed-loop system with feedforward control and scheduling function.}
% \label{fig:fig3}
% \end{center}
% \end{figure}

Specifically, at the $k$-th synchronisation cycle, once the clock offset $\hat{\theta}_i$[$k$] is measured, the local clock is adjusted by a partial amount of offset estimate, following $u_i[k] = - \alpha \hat{\theta}_i[k]$ where $\alpha $ is the coefficient of the proportional controller.

In this paper, since the frequency of clocks is assumed to be the same as nominal frequency $f_0$ (i.e., $f_i = f_0$), these do not contribute extra value to the offset. Assuming the clock period is sufficiently small, the value of clock state increment within delay duration (which is in the state dimension) is equal to the value of delays (which is in the time dimension). In other words, if clock resolution is sufficiently high, delay in the time dimension is equivalent to delay in the state dimension, and the following expressions are obtained:
\begin{equation}
P_i^{t_k + \kappa_i} - P_i^{t_k} = \kappa_i \label{eq11}
\end{equation} for the packet exchange delay and
\begin{equation}
P_i^{t_k + \kappa_i + \eta_i} - P_i^{t_k + \kappa_i} = \eta_i \label{eq12}
\end{equation} for the processing delay.

Therefore, the packet exchange delay and processing delay in the state dimension can be written in following equations
\begin{equation}
\Bigg \{ \begin{array}{ll}
\theta_i[k+1] = \theta_i[k] + u_i[k] + \omega_{\theta_i}[n] \\
\hat{\theta}_i[k] = \theta_i[k] + \kappa_i \\
u_i[k] = \alpha (0 - \hat{\theta}_i[k]) - \eta_i
\end{array}. \label{eq13}
\end{equation}
It is important to note that, ideally, the local clock offset should be corrected to target value of ($\theta^{t_k + \kappa_i} - \alpha \hat{\theta}_i[k]$) at ($t_k + \kappa_i$). However, in practice, the $i$-th clock offset can only be adjusted to ($\theta^{t_k + \kappa_i + \eta} - \alpha \hat{\theta}_i[k]$) at the time ($t_k + \kappa_i + \eta_i$), owing to the processing delay $\eta_i$ which is required for clock offset estimation. This means that the value of extra $\eta_i$ is unintentionally used to correct the local clock, this procedure is modelled as $u_i[k] = \alpha \hat{\theta}_i[k] - \eta_i$, as indicated in (13).

\begin{thm}   % use the thm environment for theorems
For any $\alpha \in (0, 2)$, the clock offset in algorithm (13) converges to ($-\bar{\kappa}_i - \bar{\eta}_i/\alpha$) for all initial conditions.
\end{thm}

\begin{pf}    % and the pf environment for proofs
The characteristic equation of the closed-loop system is
\begin{equation} \nonumber
\lambda - 1 + \alpha = 0.
\end{equation}
The eigenvalue of the characteristic equation above is
\begin{equation} \nonumber
\lambda = 1 - \alpha.
\end{equation}
As specified in \cite{Ogata:2009}, if the coefficient $\alpha$ is within the unit circle (i.e., $\alpha \in (0, 2)$), the system achieves steady state. Moreover, using standard techniques (such as $z$-transformation and final value theorem), it is easy to find that the clock offset $\theta_i$[$k$] asymptotically converges to a certain value (i.e., asymptotic error) being satisfied:
\begin{equation}
\resizebox{0.50 \textwidth}{!}
{$
\begin{split}
\begin{matrix}
\theta_i[k]\\
k\rightarrow \infty
\end{matrix} &= \lim_{z\to1}(z-1) \bigg( -\frac{z(\alpha\bar{\kappa}_i + \bar{\eta}_i)}{(z-1+\alpha)(z-1)}+ \frac{(\sigma^2_{\theta_i}-\sigma^2_{\kappa_i}-\sigma^2_{\eta_i})}{(z-1+\alpha)} \bigg) \\
&= -\bar{\kappa}_i - \frac{\bar{\eta}_i}{\alpha}.
\end{split}
$}
\end{equation} \label{equ14}
\end{pf}
Clearly, delays in the time dimension lead to the asymptotic error of steady state. However, by introducing feedforward control input $\mu_i = \bar{\eta}_i - \alpha \bar{\kappa}_i$ to the closed-loop control system (13), it is possible to obtain a zero asymptotic error. This also means that, in practice, the synchronised wireless nodes will transmit \emph{Sync} packets simultaneously, and the interference of wireless packet will occur. To avoid the collision of \emph{Sync} packet transmission, the time slot $t_{d_i}$ is allocated to the $i$-th node to guarantee only the corresponding node can access the wireless channel for \emph{Sync} packet transmission.

Finally, a more advanced closed-loop control system, with features of delay compensation and avoidance of packet collision, is written as
\begin{equation}
\Bigg \{ \begin{array}{ll}
\theta_i[k+1] = \theta_i[k] + u_i[k] + \omega_{\theta_i}[n] \\
\hat{\theta}_i[k] = \theta_i[k] + \kappa_i \\
u_i[k] = \alpha (t_{d_i} - \hat{\theta}_i[k]) - \eta_i +\mu_i
\end{array}.\label{eq15}
\end{equation}
\begin{thm}   % use the thm environment for theorems
For any $\alpha \in (0, 2)$, the offset in algorithm (15) tracks $t_{d_i}$ for all initial conditions.
\end{thm}

\begin{pf}    % and the pf environment for proofs
Using standard techniques (such as $z$-transformation), it becomes almost natural to find transfer function $G(z)$ of the closed-loop control system (15) from $t_{d_i}(z)$ to $\theta_i(z)$:
\begin{equation} \nonumber
G(z) = \frac{\theta_i(z)}{t_{d_i}(z)}=\frac{\alpha}{(z-1+\alpha)}.
\end{equation}
Therefore,
\begin{equation} \label{eq16}
G(1) = 1.
\end{equation}
This means that, by using the feedforward control, in the steady state, the clock offset $\theta_i$[$k$] always tracks desired reference input $t_{d_i}$. In other words, in the synchronised state, the $i$-th node transmits the \emph{Sync} at the allotted time slot $t_{d_i}$.
\end{pf}

\section{Simulation Results}
To validate theoretical results presented in the preceding section, simulations are conducted in Simulink employing parameters obtained from real implementation settings. In the simulations, an identical clock, which corresponds to the crystal oscillator-based clock on Node $1$ in Fig. 1, is simulated. The clock offset is subject to a random perturbation with variance of $244.4990\times10^{-12}$, which is taken from \cite{Zong:2020a}. The synchronisation cycle is configured to $1s$. In addition, the mean values of packet exchange delay and processing delay are set to $349 \mu s$ and $514 \mu s$, respectively (see \cite{Zong:2020b}). The network consisting of one master and a single slave node is used. The configurations of simulations are summarised in Table 1.

\begin{table}[hb]
\begin{center}
\caption{Simulation configurations}\label{tb:margins}
\begin{tabular}{ccc}
Symbol & Value & Unit \\\hline
$\theta_i[0]$ & 0.6 & Second ($s$) \\
$\sigma_{\theta_i}^2$ & $244.4990\times10^{-12}$ &  \\
$T$ & $1$ & Second ($s$) \\
$\bar{\kappa}_i$ & $349$ & Microsecond ($\mu s$) \\
$\bar{\eta}_i$ & $514$ & Microsecond ($\mu s$) \\
$\alpha$ & $0.5$ & \\
$t_{d_i}$ & $0, 9.15$ & Microsecond ($\mu s$) \\ \hline
\end{tabular}
\end{center}
\end{table}

Fig. 3 shows the evolution of clock offset under the closed-loop control system (13). Clearly, it is demonstrated that, without the aid of feedforward control, the clock offset converges to an asymptotic error $ -\bar{\kappa}_i -\bar{\eta}_i/\alpha  = -1.212 ms$, after several synchronisation cycles.

Moreover, Fig. 4 indicates that, by using feedforward control and non-zero reference input, in the steady state, the clock offset tracks reference input $t_{d_i} = 9.15ms$. This means that, in practice, the $i$-th slave node transmits the \emph{Sync} at its allotted time slot $t_{d_i}$ within each synchronisation cycle.

\begin{figure}
\begin{center}
\includegraphics[width=8.4cm]{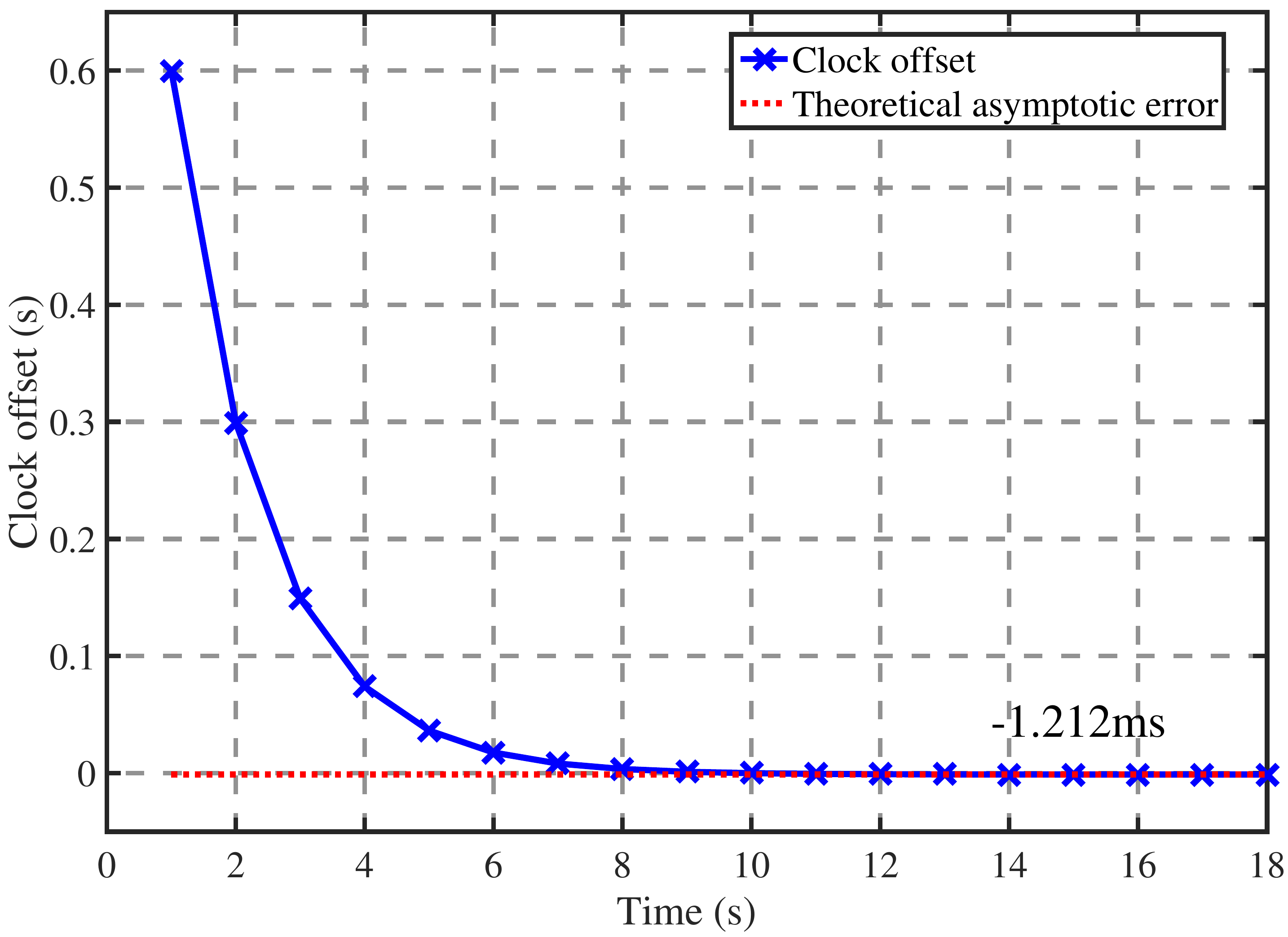}    % The printed column width is 8.4 cm.
\caption{Evolution of clock offset on the closed-loop control system without delay compensation strategy.}
\label{fig:fig4}
\end{center}
\end{figure}

\begin{figure}
\begin{center}
\includegraphics[width=8.4cm]{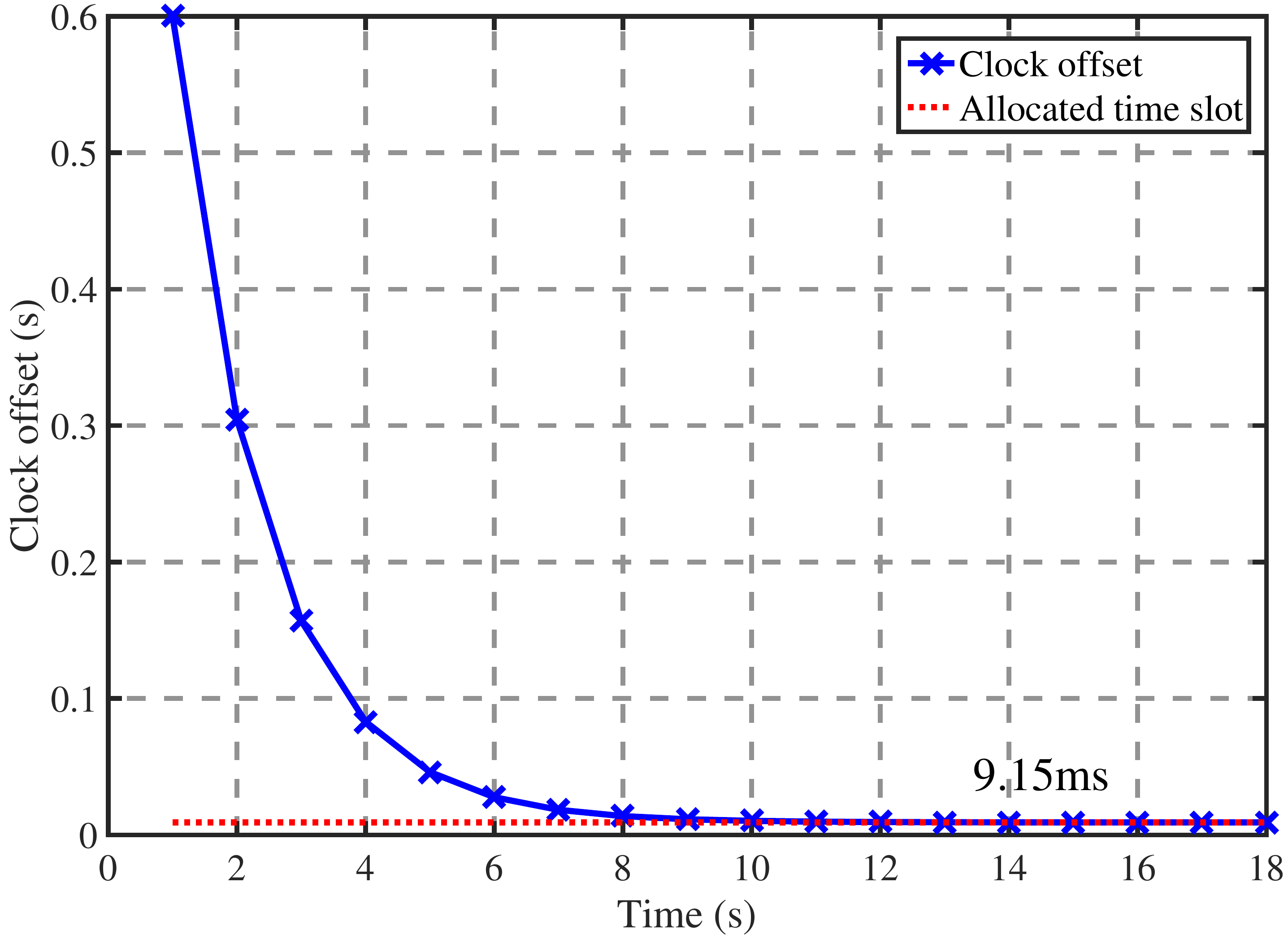}    % The printed column width is 8.4 cm.
\caption{Evolution of clock offset on the closed-loop control system with delay compensation and avoidance of packet collision strategies.}
\label{fig:fig5}
\end{center}
\end{figure}

\section{Experimental Results}
To evaluate the performance of the proposed PkCOs synchronisation scheme with features of delay compensation and packet collision avoidance, it is implemented on the Atmel SMART SAM R21 board developed in \cite{Zong:2018b}. The experiments consider a two-node wireless network consisting of one master node and a single slave node. The master node is connected to a Trimble ThunderBolt E GPS Disciplined Clock for providing the PPS (Pulse Per Second) signal. Once it receives the PPS signal from GPS, the master issues an external interrupt to toggle the GPIO PA19 pin firstly. Meanwhile, it broadcasts a \emph{Sync} packet to the wireless channel directly.

As for the slave node, the PkCOs state is represented by a 32-bit COUNT register of the Real-Time Clock (RTC) module, which uses a $32.768$kHz external crystal oscillator as the clock source. When the COUNT register matches 32767 of the compare register COMP, it is reset to zero. At the same time, an RTC interrupt is triggered, and then, a new cycle begins. In the RTC interrupt handler, the GPIO PA19 pin is toggled first; then, a wireless \emph{Sync} packet is transmitted to the channel directly.

In the slave's receive mode, once addressing fields of the received \emph{Sync} packet match local addresses, an AMI (Address Match Interrupt) is issued to generate a timestamp by reading the COUNT register. After the timestamp is generated, the local clock is corrected by using estimated clock offset as described in (7).

The signal outputs from master and slave nodes are attached to a logic analyser that records the time of master's \emph{Sync} packet transmission and slave clock's firing. Also, the difference between \emph{Sync} packet transmission time and signal toggled in the RTC handler of slave node is calculated for analysing synchronisation performance. To guarantee a fair comparison of performance, the precision in the synchronised state, denoted by $\Delta_i$, is defined as the difference among the $i$-th clock fires time $t_{P_i=\varphi_i}$, allocated time slot $t_{d_i}$ and perfect clock fires time $t_{P_i=\varphi_0}$ of the master node, chosen as the reference. That is
\begin{equation} \label{eq17}
\Delta_i := t_{P_i = \varphi_0} - t_{d_i} - t_{P_i = \varphi_i}.
\end{equation}
Many time synchronisation protocols are proposed and implemented in several hardware platforms that select different types of crystal oscillator as clock sources; thus, the direct comparison of these results is unfair, and obtained conclusions are also questionable (\cite{Masood:2017}). In addition to $\Delta_i$, we also adopt the clock tick as an evaluation metric, in order to compare the performance of multiple synchronisation algorithms in different hardware environments. This metric relates to maximum achievable accuracy on the underlying platform, leading to reasonable results.

Fig. 5 presents the evolution of synchronisation precision over time. Time is measured in terms of the number of synchronisation cycles. The precision converges to some extent to a steady state with an asymptotic error, which is close to the theoretical asymptotic error of ($ \bar{\kappa}_i + \bar{\eta}_i/\alpha = 1.1895 ms$). However, due to the limitation of clock resolution (i.e., the clock period $\tau_0 = 30.5 \mu s$), the assumption of (11), (12) cannot be met, and the value of delays (i.e., $ \bar{\kappa}_i + \bar{\eta}_i/\alpha $) is impossible to be integer multiple of the clock period. Thus, there always exists a bias between the value for compensating delays, saying $26.6 \mu s$.

\begin{figure}
\begin{center}
\includegraphics[width=8.4cm]{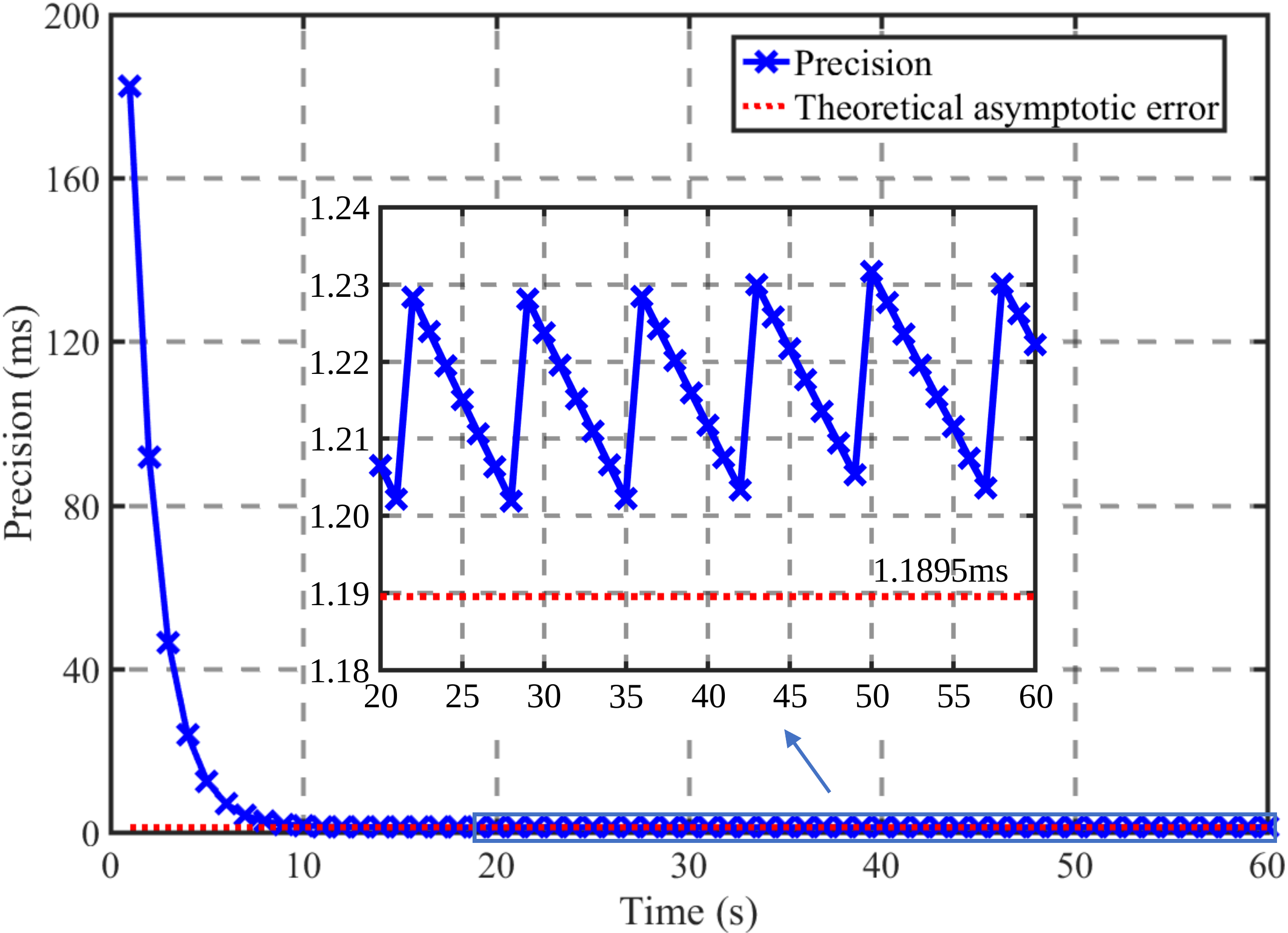}    % The printed column width is 8.4 cm.
\caption{Evolution of synchronisation precision $\Delta_i$ under the PkCOs scheme without delay compensation ($\alpha = 0.5, \bar{\kappa}_i = 518.5 \mu s, \bar{\eta}_i = 335.5 \mu s$).}
\label{fig:fig6}
\end{center}
\end{figure}

Moreover, in the experiments, only the clock offset is corrected, and the local clock's frequency is slightly different from that of the GPS clock. The existence of clock skew leads to sawtooth behaviour of achieved synchronisation precision. In future work, the clock frequency correction scheme will be applied to local clocks, in order to further improve synchronisation performance.

\begin{figure}
\begin{center}
\includegraphics[width=8.4cm]{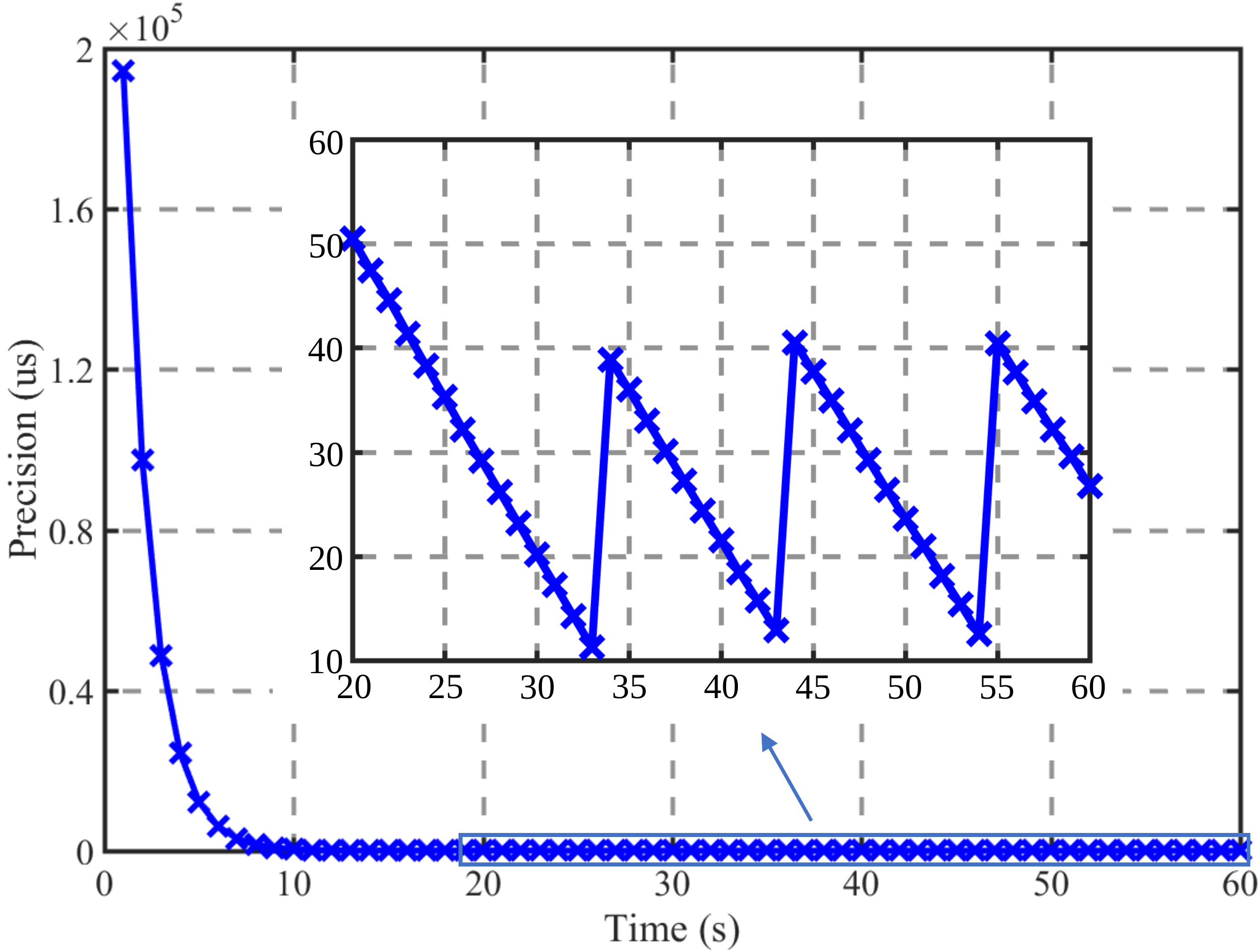}    % The printed column width is 8.4 cm.
\caption{Evolution of synchronisation precision $\Delta_i$ under the PkCOs protocol with delay compensation and avoidance of packet collision ($\alpha = 0.5, \bar{\kappa}_i = 518.5 \mu s, \bar{\eta}_i = 335.5 \mu s$).}
\label{fig:fig7}
\end{center}
\end{figure}

In Fig. 6, by employing feedforward control $\mu_i$ to packet coupling scheme, the achieved synchronisation accuracy improves to $26.3 \mu s$ ($1$ ticks). Even though \cite{Degesys:2007} utilises MAC-level timestamping mechanism to generate timestamp, the achieved synchronisation is only around $100 \mu s$ ($3$ ticks). In addition, \cite{Elson:2002} reports that RBS is capable of achieving synchronisation with the precision of about $11 \mu s$ ($11$ ticks). By using similar hardware configurations, the proposed synchronisation method has a clear potential to overcome the results in \cite{Elson:2002} and \cite{Degesys:2007}.

\begin{figure}
\begin{center}
\includegraphics[width=8.4cm]{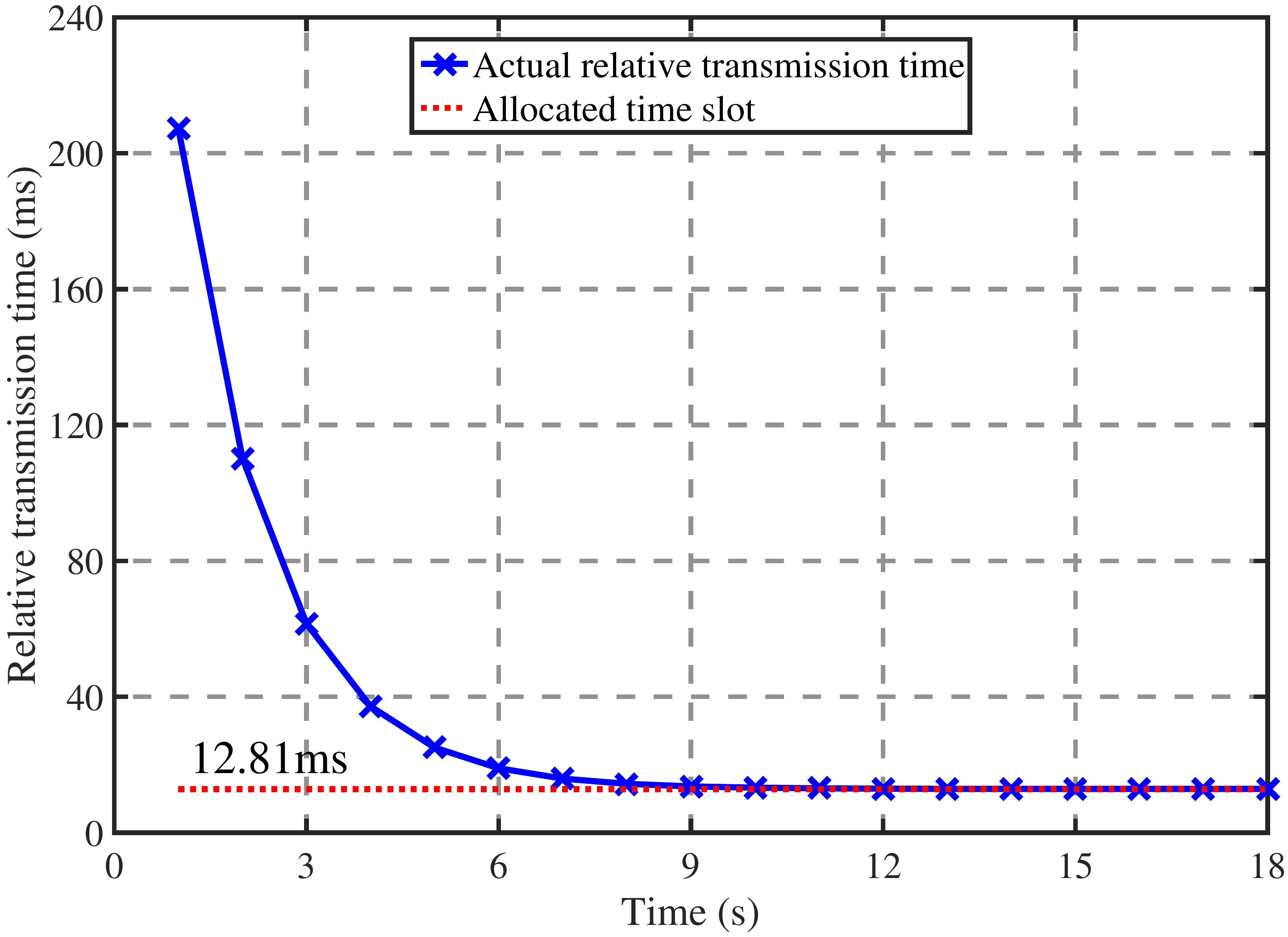}    % The printed column width is 8.4 cm.
\caption{Transmission time relative to the master node during each synchronisation cycle ($t_{d_i} = 12.81ms$).}
\label{fig:fig8}
\end{center}
\end{figure}

Fig. 7 demonstrates results of the utilisation of feedforward control and non-zero reference input $t_{d_i} = 12.81ms$ in the system. Clearly, in the steady state (i.e., synchronised state), the time of the $i$-th clock packet transmission event relative to the master node is $12.81ms$. In other words, the $i$-th node's clock always tracks reference input $t_{d_i}$. This means that the proposed packet coupling scheme can naturally allocate the time slot for \emph{Sync} packet transmission, thereby minimising the possibility of packet collision.

\section{Conclusion}
In this paper, through mathematical modelling and experimental validation of packet exchange delay and processing delay, the impacts of two delays on PkCOs are studied. The analysis includes packet exchange delay and processing delay in both time and state dimensions. The feedforward control (i.e., delay compensation strategy) is introduced in the packet coupling scheme to cancel delays' effects. Meanwhile, the proposed simple packet coupling scheme allows the natural allocation of a time slot for \emph{Sync} packet transmission. The rigorous theoretical proofs validate the effectiveness of the proposed PkCOs scheme with delay compensation and packet avoidance characteristics. Finally, the experimental results show that, in the PkCOs synchronisation protocol, the utilisation of feedforward control can achieve synchronisation with the precision of $26.3 \mu s$ ($1$ tick). It is also capable of allocating the \emph{Sync} packet to a specified time slot, thereby minimising the possibility of packet collision in wireless sensor networks.

In future work, the hardware resource consumption of PkCOs requires to be analysed in the experiments. Moreover, it is also necessary to study synchronisation precision of the PkCOs protocol in multi-hop large-scale WSNs.

\end{document}